\RequirePackage{ifpdf}
\ifpdf 
\documentclass[pdftex]{sigma}
\else
\documentclass{sigma}
\fi

\numberwithin{equation}{section}

\begin{document}

\newcommand{\Tr}{\mathop{\mathrm{Tr}}\nolimits}
\newcommand{\IIm}{\mathop{\mathrm{Im}}\nolimits}
\newcommand{\pd}[2]{\frac{\partial #1}{\partial #2}}
\newcommand{\od}[2]{\frac{d#1}{d#2}}
\newcommand{\csch}{\mathop{\mathrm{csch}}\nolimits}
\newcommand{\floor}[1]{{\left\lfloor #1 \right\rfloor}}
\newcommand{\gfatop}[2]{{\genfrac{}{}{0pt}{1}{#1}{#2}}}

\allowdisplaybreaks

\renewcommand{\PaperNumber}{094}

\FirstPageHeading

\renewcommand{\thefootnote}{$\star$}

\ShortArticleName{Vacuum Energy as Spectral Geometry}

\ArticleName{Vacuum Energy as Spectral Geometry\footnote{This paper is a
contribution to the Proceedings of the 2007 Midwest
Geometry Conference in honor of Thomas~P.\ Branson. The full collection is available at
\href{http://www.emis.de/journals/SIGMA/MGC2007.html}{http://www.emis.de/journals/SIGMA/MGC2007.html}}}

\Author{Stephen A. FULLING}

\AuthorNameForHeading{S.A. Fulling}

\Address{Department of Mathematics, Texas A\&M University,
 College Station, Texas, 77843-3368, USA}

\Email{\href{fulling@math.tamu.edu}{fulling@math.tamu.edu}}

\URLaddress{\url{http://www.math.tamu.edu/~fulling/}}

\ArticleDates{Received June 21, 2007, in f\/inal form September 14, 2007; Published online September 26, 2007}

\Abstract{Quantum vacuum energy (Casimir energy) is reviewed for a
mathematical audien\-ce as a topic in spectral theory.
 Then some one-dimensional systems are solved exactly, in terms of closed
classical paths and periodic orbits.
 The relations among local spectral densities, energy densities, global
 eigenvalue densities, and total energies are demonstrated.
This material provides background and motivation for the treatment of
 higher-dimensional systems (self-adjoint second-order partial
dif\/ferential operators) by semiclassical approximation and other methods.}


\Classification{34B27; 81Q10; 58J50}

\section{Introduction} \label{sec:intro}

 Vacuum energy is a concept arising in quantum f\/ield theory, with
observable consequences of considerable current interest in
physics \cite{casimir,PMG,milton,BMM,ILMC}.
 Here, however, I treat it
 as a purely mathematical problem, an underdeveloped aspect of the
spectral theory of self-adjoint second-order dif\/ferential
operators.

 Vacuum energy is related to the oscillatory terms in the spectral
density, which are associated with the periodic orbits in the
classical mechanics determined by the operator as Hamiltonian
\cite{BB3,BB2',colin,gutz4,DG,CRR,BB}.
 The information concerning these oscillations is lost from the
much-studied short-time expansion of the heat kernel
 \cite{minak,kac,greiner,gilkey,BB,kirsten},  but
some of it remains in the asymptotics of another integral kernel,
the cylinder kernel, which includes the vacuum energy directly
\cite{lukosz2,BH,hays,FG,systemat,funorman}.
The local density of vacuum  energy is related analogously to a local
spectral density, associated with classical paths that are closed
but not necessarily periodic.
 Thus the vacuum energy and its density are  probes of the detailed
geometry associated with the operator.

 The connection between closed classical paths and Casimir energy
 is implicit in the two papers of Balian and Duplantier \cite{BD},
which treat those two aspects of the electromagnetic f\/ield
separately.
 In recent years it has become a serious tool for the calculation
of vacuum energies
\cite{JR,SS,MSSvS,funorman,SJ,schaden1,schaden2,HJKS}.
 In one spatial dimension it reduces to the classical method of
images, which has been applied to Casimir calculations since at
least 1969 \cite{BmC}.

 The main purpose of this article is to demonstrate in detail, for
the simplest one-dimensional models,
  the connections among closed orbits, spectral densities, and
  vacuum energy densities, as well as the corresponding global
  quantities.
 Sections \ref{sec:twist} and \ref{sec:bdry} are based on research
notes that I have used for some years with my students and
collaborators; the fact that they have not been published and
available for citation has become increasingly inconvenient.
 First, however, in Section \ref{sec:general} I present the basics
of vacuum energy in a more general context to a mathematical
audience.
 This material is based on talks given at the Workshop on
Semiclassical Approximation and Vacuum Energy at Texas A\&M
University in January, 2005, and the Workshop on Spectral Theory
and Its Applications
 at the Isaac Newton Institute, University of Cambridge, in July,
2006.
  I expect to publish elsewhere \cite{ini} a more complete
discussion of the physics of the subject, still for a mathematical
audience, along with a review of the recent work on vacuum energy
in quantum graphs \cite{wilson}.

 \section{Vacuum energy (and energy density) in general}
  \label{sec:general}

 \subsection{Spectral theory}
  Let $H$ be a second-order, elliptic,  self-adjoint
 partial dif\/ferential operator, on~scalar functions,
 in a $d$-dimensional region $\Omega$.
The prototype situation is a \emph{billiard}:
 \[  H=-\nabla^2, \qquad \Omega\subset\mathbf{R}^d, \]
with  boundary conditions that make the operator self-adjoint
 (such as the Dirichlet condition, $u=0$ on $\partial\Omega$).
The treatment can be generalized to the
electromagnetic f\/ield (vector functions)~-- which is the case of
greatest physical interest~--
and to other boundary conditions,
to Riemannian manifolds (Laplace--Beltrami operators),
  potentials  ($H= -\nabla^2 + V(x)$), applied magnetic f\/ields
(gauge-invariant Laplacians on vector bundles), etc.
For simplicity, assume that the spectrum of $H$ is nonnegative and
that if $0$ is  an eigenvalue, then the eigenfunction is constant.
 Of course, for precise theorems some smoothness hypotheses on the
boundary, potential, etc.\ are needed.

  For the moment let us assume that the spectrum is discrete, as
  will be so if $\Omega$ is a compact billiard. In this case a
  f\/inite total vacuum energy is expected.
We review the spectral decomposition and functional calculus.
Let $\varphi_n$ be the normalized eigenfunctions:
\[ H\varphi _n= \lambda_n \varphi _n , \qquad
 \|\varphi _n\|^2 = \int_\Omega |\varphi _n(x)|^2\, dx =1. \]
Def\/ine $\omega_n = \sqrt{\lambda_n}$.
 Functions of the operator $H$ are def\/ined by
 \[ f(H)u \equiv \sum_{n=1}^\infty f(\lambda_n)
  \langle\varphi _n, u\rangle \varphi _n, \qquad
 \langle\varphi_n,u\rangle \equiv \int_\Omega \overline{\varphi_n(x)}
 u(x)\,dx. \]
 At least formally, $f(H)$ is given by an integral kernel:
\[ f(H)u(x) = \int_\Omega G(x,y)u(y)\, dy,
 \qquad
 G(x,y) = \sum_{n=1}^\infty f(\lambda_n) \varphi _n(x)
 \overline{\varphi _n(y)} .\]
 In general $G$ is a distribution, but
if $f$ is suf\/f\/iciently rapidly decreasing, $G$ is a
smooth function, and the trace is def\/ined:
  \[ \Tr G \equiv\int_\Omega G(x,x)\,dx = \sum_{n=1}^\infty
   f(\lambda_n) \equiv \Tr f(H). \]

 The prototype is the \emph{heat kernel}, $G(x,y) = K(t,x,y)$,
  corresponding to the parametrized function
 $f_t(\lambda)= e^{-t\lambda}$.  Then $u(t,x) = f_t(H) u_0$ solves
 \[ \pd {u}{t} = -Hu,   \qquad u(0,x) = u_0(x). \]
 It is well known \cite{greiner,BG,gilkey,kirsten} that $K$ for a
billiard
 has the asymptotic expansion
\begin{gather} \Tr K = \sum_{n=1}^\infty e^{-t\lambda_n} \sim
\sum_{s=0}^\infty b_s t^{(-d+s)/2},
 \label{heattr} \end{gather}
where each term has a global geometrical signif\/icance~-- for
instance, $b_0$ is proportional to the volume of~$\Omega$.
 The inverse Laplace transform of the leading term in $\,\Tr K$
  gives the leading
behavior at large $\lambda$ of the density of eigenvalues,
 and the higher-order terms correspond similarly to lower-order
corrections to the eigenvalue distribution, \emph{on the average}
 \cite{brownell,hormander,lgacee1}.
 (When $V(x)$ is a conf\/ining potential \cite{gutz4,BB2',BB},
  $H = -\nabla^2 +V$ may have
discrete spectrum even though its spatial domain, $\Omega$, is not
compact.  In such a case the form of the asymptotic expansion of
$\,\Tr K$, and the resulting asymptotic expansion of the eigenvalue
distribution, may be rather dif\/ferent.
 For example, the eigenvalues of a one-dimensional harmonic oscillator
 are evenly spaced in $\lambda$, whereas those of a one-dimensional
billiard are evenly spaced in $\omega$.)

The \emph{cylinder kernel} (also called \emph{Poisson kernel}),
 $T(t,x,y)$,
corresponds to the parametrized function
  $f_t(\lambda) = e^{-t\sqrt{\lambda}}$. That is,
 $f_t(H) u_0$   is the solution of
  \[ \pd {^2u}{t^2} = Hu,   \qquad u(0,x) = u_0(x), \]
 that is well-behaved as $t\to+\infty$.
   ($T$ is a boundary value of a derivative of the fundamental
solution of the elliptic operator
 $H- \pd{^2}{t^2} $ in $\Omega\times\mathbf{R}$.
 It is the (``imaginary-time'') analytic conti\-nuation of the
 time derivative of the \emph{Wightman function},
 a certain fundamental solution of the hyperbolic operator
 $H+\pd{^2}{t^2}\, $.)
We have
 \[ T(t,x,y) = \sum_{n=1}^\infty e^{-t\omega_n} \varphi _n(x)
\overline {\varphi _n(y)} , \qquad
 \Tr T=\int_\Omega T(t,x,x) \,dx = \sum_{n=1}^\infty e^{-
t\omega_n}. \]
As $t\downarrow0$ one has the asymptotics \cite{GG,FG,BM}
 \begin{gather}  \Tr T \sim
\sum_{s=0}^\infty e_s t^{-d+s}
+\sum^\infty_\gfatop{s=d+1}
{s-d \text{ odd}} f_s t^{-d+s} \ln t,
 \label{cyltr}\end{gather}
as described in the following theorem.
 (For a more complete, but succinct, statement of the connection
with Riesz means, see \cite{funorman}.  See also
  \cite{CVZ,E,EF,systemat}.)

 \begin{theorem} \label{thm:Rieszconn}
 The traces of the heat kernel and the cylinder kernel of a
positive, self-adjoint, second-order linear differential operator
in dimension~$d$ have the asymptotic expansions \eqref{heattr} and
\eqref{cyltr}, and precisely parallel expansions hold for the
local (untraced) diagonal values of those kernels.
 The $b_s$ are proportional to coefficients in the high-frequency
asymptotics of Riesz means of  $N$ (or~$P$) with respect to $\lambda$.
 The $e_s$ and $f_s$ are proportional to coefficients in the
asymptotics of Riesz means  with respect to $\omega$.
 If $d-s$ is even or positive,
\[e_s= \pi^{-1/2} 2^{d-s} \Gamma((d-s+1)/2) b_s. \]
 If $d-s$ is odd and negative,
\[ f_{s} = \frac{(-1)^{(s-d+1)/2}2^{d-s+1}}
 {\sqrt{\pi}\, \Gamma((s-d+1)/2)} \, b_s, \]
but $e_{s}$ is undetermined by the $b_{r}$.
\end{theorem}

\subsection{Vacuum energy}
We can now  def\/ine the \emph{vacuum energy} as
 the coef\/f\/icient with $d-s=-1$,
\begin{gather}
  E \equiv - {\textstyle \frac12} e_{1+d}.
 \label{energy}\end{gather}
 Formally, $E$ is the ``f\/inite part'' of
 \[ \frac12 \sum_{n=1}^\infty \omega_n =
   \left . -\,\frac12\,\od{}{t}\sum_n e^{-\omega_n t}
 \right |_{t=0}. \]
(When $f_{1+d}\ne0$, $E$ is actually def\/ined only modulo a
multiple of $f_{1+d}\,$, because of the scale ambiguity
in the argument of the logarithm. That complication will not
arise in the problems studied in this paper.)

 The prototype example  is
 \begin{gather}
  \Omega= (0,L), \qquad
  H = -\,\od{^2}{x^2}; \qquad
 \omega_n = \frac{n\pi}{L}, \qquad
 \varphi _n(x) = \sin \left (\frac{n\pi x} L\right ).
 \label{interval} \end{gather}
One can evaluate the cylinder kernel directly from the
spectral decomposition as
 \[  T(t,x,y) = \frac2L \sum_{n=1}^\infty
\sin\left (\frac{n\pi x}L\right) \sin \left(\frac{n\pi y}L\right)
  e^{-(n\pi/L)^2 t},\]
 or by the method of images as a sum over classical paths,
 \begin{gather}
  T(t,x,y)  = \frac t\pi \sum_{N=-\infty}^\infty
\left[  \frac1{(x-y-2NL)^2 + t^2} - \frac1{(x+y-2NL)^2 + t^2}
\right].
 \label{intimage}\end{gather}
  Either sum can be evaluated in closed form as
\begin{gather}
  T(t,x,y) =\frac1{2L}
 \left[\frac{\sinh (\pi t/L)}{\cosh(\pi t/L)
  - \cos\bigl(\pi(x-y)/L\bigr)}
 -  \frac{\sinh (\pi t/L)}{\cosh(\pi t/L) - \cos\bigl(\pi(x+y)/L\bigr)}
 \right] .
 \label{intcyl} \end{gather}
It follows that
 \[ \Tr T
= \frac12\, \frac{\sinh(\pi t/L)}{\cosh(\pi t/L) -1}  -\frac12
 \sim \frac L{\pi t}  - \frac12
+\frac {\pi t}{12 L} +O(t^3) . \]
Thus the energy,  the $O(t)$ term times $-\frac12$, is
\begin{gather}
  E = -\,\frac{\pi}{24 L}.
 \label{inten}\end{gather}
This formula has been known for  many years (e.g., \cite{BH}).

 Another simple example, or class of them,
  involves a vector bundle over
the circle, coordinatized as  $\Omega= (0,L)$ \cite{isham,FGR}.
Again $H = -d^2/dx^2$, but now the f\/ield is ``twisted'' so~that
 \begin{gather}
  u(L) = e^{i\theta} u(0), \qquad u'(L) = e^{i\theta} u'(0).
 \label{twisbc}\end{gather}
 The eigenfunctions and eigenvalues are
\begin{gather}   \varphi_n(x) = e^{i(2\pi n+\theta)x/L)}
 \quad (n\in\mathbf{Z}), \qquad
\omega_j^\pm = \frac{2\pi j \pm \theta} L
 \quad(j\in  \mathbf{N} \text{ or } \mathbf{Z}^+).
 \label{twiseig}\end{gather}
 One can then show (see Section~\ref{sec:twist}) that
 \begin{gather}
 E_\theta = -\,\frac{\pi}L\, B_2\left (\frac\theta{2\pi}\right ) =
 -\,\frac\pi{12L}\left [ 2 - 6\,\frac{\theta}{\pi}
 +3 \left (\frac\theta\pi\right )^2\right ].
 \label{twisen} \end{gather}
 ($B_2$ is a Bernoulli polynomial.)
 What is intriguing about this example is its dependence on the
parameter~$\theta$; $E_\theta$ can be positive as well as negative.
Its extreme values are
  $ E_0 = -\, \frac\pi{6L}$ and $E_\pi = +\, \frac\pi{12L}\,$;
these are the cases where the two eigenvalue sequences coincide,
  so the gaps in the spectrum are largest.
$E_\theta$ passes through $0$ for $\theta\approx0.42$,
which is close to the point $\pi/2$ where the eigenvalues are
equally spaced.
We observe that
 for an individual eigenvalue sequence, the sign and
magnitude
  of its contribution to the energy are determined by the \emph{phase} of
  the spectral oscillation. Here that phase is controlled by $\theta$;
  in higher-dimensional systems the phase of the oscillations
  associated with
 a periodic classical orbit is controlled by the famous
\emph{[Kramers--Morse--Keller--Gutzwiller--]Maslov index}
 (see, e.g., \cite{littlejohn}),
 whose signif\/icance for vacuum energy
has only begun to be explored \cite{schaden1,schaden2,EFKKLM}.

\subsection{Vacuum energy density}
 For present purposes the  energy \emph{density} can be def\/ined
simply by leaving out the integration over $x$ in the cylinder
trace and proceeding as before.
 Let  $P(\lambda,x,y)$ be the integral kernel of the orthogonal projection
 onto the part of $L^2(\Omega)$
 corresponding to spectrum less than or equal to $\lambda$;
 since
 \[K(t,x,y) = \int_0^\infty e^{-t\lambda}\, dP(\lambda,x,y), \]
 we can also def\/ine $P$ as the inverse Laplace transform of the
heat kernel
 (the \emph{exact} heat kernel, not its small-$t$ series).
Then
 \[T(t,x,y) =  \int_0^\infty e^{-t\sqrt \lambda} \,dP(\lambda,x,y) ,\]
 and, as mentioned in Theorem \ref{thm:Rieszconn},
  its trace has an expansion completely analogous to
\eqref{cyltr} with $x$-dependent coef\/f\/icients:
\[T(t,x,x) \sim
\sum_{s=0}^\infty e_s(x) t^{-d+s}
+ \sum^\infty_\gfatop{s=d+1}{s-d \text{ odd}}
  f_s(x) t^{-d+s} \ln t.\]
 We can now def\/ine the energy density as
 \[E(x) = - {\textstyle \frac12} e_{1+d}(x) .\]
 In the quantum f\/ield theory, $E(x)$
 (known in that context as $T_{00}(x)$ def\/ined with $\xi=\frac14$)
is formally the f\/inite part of
 \[\frac12 \left [\left (\pd ut\right )^2 + u\,Hu\right ]. \]

  When the spectrum is discrete,
 \[P(\lambda,x,y) = \sum_{\lambda_n\le\lambda} \varphi_n(x)
\overline{\varphi_n(y)}, \]
 and the integral of $P(\lambda,x,x)$ over $\Omega$ equals
$N(\lambda)$, the number of eigenvalues less than or equal
to~$\lambda$.
But one reason for the importance of the energy density is that it
remains meaningful when $H$ has some continuous spectrum.
 If the spectrum is absolutely continuous,
 \[\sigma(\omega,x) \equiv \od{}{\omega}P(\omega^2,x,x) \]
 exists (as a Radon--Nikodym derivative) and constitutes a
 \emph{local spectral density}.
  (It is a density in two senses~-- with respect to both $\omega$
  and~$x$.)

Here, also, there is a prototypical example,
  the  half-line with a Dirichlet endpoint:
  \begin{gather}
 \Omega = (0,\infty),\qquad H= -\,\od{^2}{x^2}, \qquad
 u(0)=0.
 \label{hfline}\end{gather}
 The eigenfunction expansion is the Fourier sine transform,
equivalent to the projector
\[P(\lambda,x,y) =  \int_0^{\sqrt{\lambda}} \frac2{\pi}\,
  \sin (kx) \sin(ky)\,dk. \]
 The solution for the cylinder kernel by the method of images is
now a simpler analogue of \eqref{intimage}:
\begin{gather}
  T(t,x,y) =\frac{t}{\pi} \left[\frac1{(x-y)^2+t^2} - \frac1{(x+y)^2 +
t^2}\right].
 \label{hflinecyl}\end{gather}
 Thus
\begin{gather}
  T(t,x,x) \sim
 \frac1{\pi t} -    \frac{t}{\pi(2x)^2}
\sum\limits^\infty_{k=0} (-1)^k \left(\frac{t}{2x}\right)^{2k}
\quad\text{as $t\downarrow 0$},
 \label{hflinediag}\end{gather}
so
 \begin{gather} E(x) = \frac1{8\pi x^2}.
\label{hflineen} \end{gather}

 Recall
  that in any one-dimensional billiard the untraced diagonal
value of the heat kernel is simply
 \[K(t,x,x) \sim (4\pi t)^{-d/2}  + O(t^\infty) \]
(for f\/ixed $x$ in the interior of $\Omega$)
  regardless of boundary conditions.
In contrast, we see from~\eqref{intcyl}, \eqref{twisen}, and
\eqref{hflinediag}
 that $T(t,x,x)$  is sensitive to the \emph{global} geometry~--
 the length of the interval, the nature of the boundary conditions,
the structure (angle~$\theta$) of the vector bundle.
 (In case~\eqref{twisen}, $E(x)$ is independent of~$x$ and
equals  $E/L$.)
Therefore, $T(t,x,x)$, $E(x)$, and $E$ are interesting objects to
study from the point of view of inverse problems and other aspects
of spectral geometry.

\begin{figure}[t]
 \begin{minipage}[t]{78mm}\centering
 \includegraphics[width=75mm]{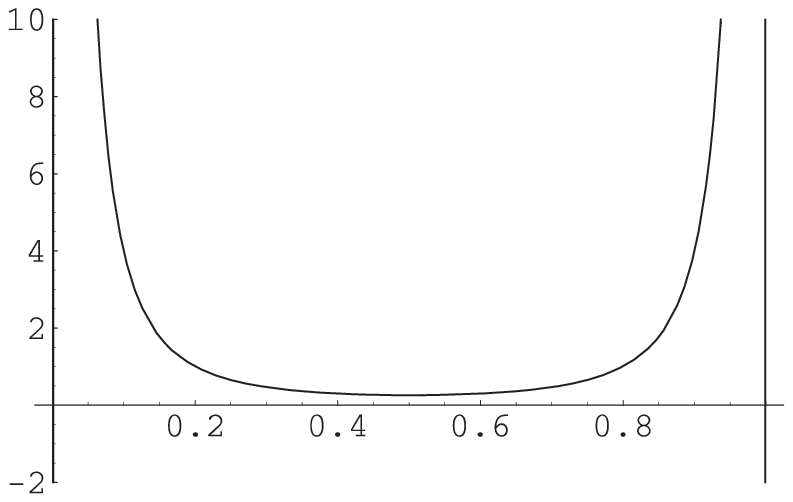}
  \caption{Boundary energy density for $\Omega=(0,1)$.}
  \label{fig:int}
  \end{minipage} \hfill
 \begin{minipage}[t]{75mm}\centering
\includegraphics[width=75mm]{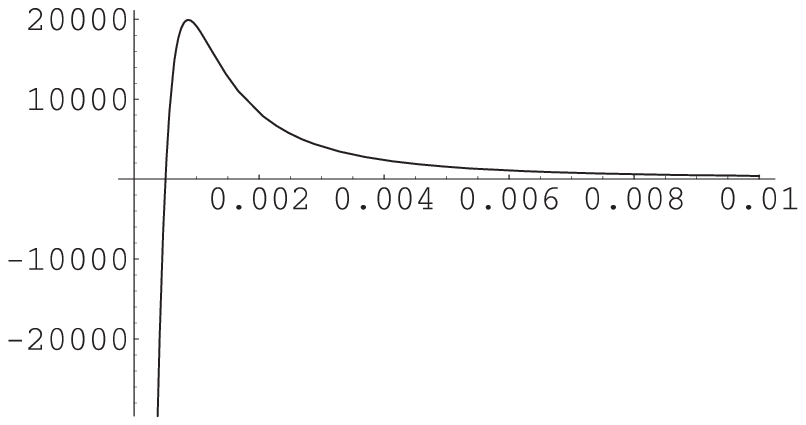}
\caption{Regularized energy density $E(t,x)$ for
 $\Omega=(0,\infty)$ and $t=0.001$.}
\label{fig:hfline}
\end{minipage}
\end{figure}

For the f\/inite interval \eqref{interval}, one f\/inds
\begin{gather} E(x)=  -\,\frac\pi{24L^2}
 +\frac\pi{8L^2} \,\csc^2\left (\frac{\pi x}{L}\right ) .
\label{intenden}\end{gather}
 This function is graphed in Fig.~\ref{fig:int}.
  Now
\[\frac\pi{8L^2} \,\csc^2\left (\frac{\pi x}{L}\right ) \sim
 \frac1{8\pi x^2} \text{ as $x\to0$}, \]
 with a similar expansion  as $x\to L$.
 Therefore, $E(x)$ displays the boundary behavior \eqref{hflineen}
in addition to a spatially homogeneous term.
 But now notice something strange:
 The ``total'' energy $E$, \eqref{inten}, is equal to the integral
over $\Omega$ of the spatially homogeneous term only, while the
integral of the boundary terms diverges!
 In physical terms, \emph{the renormalized energy is not equal to the
integral of the renormalized energy density.}
 The physical signif\/icance of this fact is beyond the scope of this
article.  (In fact, it is still controversial.)
 Mathematically, however, it is just an instance of
 nonuniform  convergence:  The limit $t\to0$ cannot be interchanged
with the limit $x\to0$, and hence with the integration over~$x$.

To examine this phenomenon more closely, it is convenient to
return to the half-line \eqref{hfline}.
 Since the latter is spatially inf\/inite, one would not necessarily
expect its total energy to be def\/ined; however, we notice that the
integral of \eqref{hflineen} converges at inf\/inity.
 Let us keep $t$ positive (but small) and consider the
 \emph{regularized}  vacuum energy density
 \[E(t,x) = -\,\frac12\, \pd{}t T(t,x,x)- \frac1{\pi t^2}=
     -\, \frac1{2\pi} \, \frac{t^2-4x^2}{(t^2+4x^2)^2} . \]
(See Fig.~\ref{fig:hfline}, noting the compressed vertical scale.
 The subtracted term is the contribution of the f\/irst term in
 \eqref{hflinecyl}
 (or the f\/irst term in \eqref{hflinediag}),
  identif\/ied with the ubiquitous but unobservable vacuum energy of
 inf\/inite empty space.
 Only the second term in \eqref{hflinecyl} is of interest.)
 The limit $t\to0$ reproduces the renormalized energy density
\eqref{hflineen},
 which would imply an inf\/inite total energy if integrated.
 However,
 \[ E(t) \equiv \int_0^\infty E(t,x)\,dx =0 \quad\text{for all $t>0$},
\]
which implies a vanishing total energy if one now takes $t\to0$.
 This \emph{disappearing divergence} is a general property of
energy densities that behave like $x^{-2}$ near a boundary.
 (It has been conf\/irmed for the corners of a rectangle or rectangular
                      parallelepiped \cite{EFKKLM}.)
 By dimensional analysis any correlate of such a term in $E(t)$
 must be proportional to $ t^{-1}$ and hence must come from a term
 $f_0\ln t$ in \eqref{cyltr},
 but no such term can exist.
 (It would necessarily match a $\,\ln t$ term in the heat
 kernel~\cite{CVZ,GG,BGH,systemat}, which does
not exist for the class of operators considered here.)
 For the f\/inite interval, it will be shown in Section~\ref{sec:bdry} that
 the regularized total energy does not vanish, but does approach
the renormalized energy \eqref{inten} as $t\to0$.

\subsection{But what about the zeta function?}

 Many mathematically inclined researchers on vacuum energy have
def\/ined it in terms of  \emph{zeta functions}.
For example, \eqref{inten} is often expressed as
  \[ \frac12\sum_{n=1}^\infty \frac{n\pi}{L} \mathrel{``{=}"}
 \frac{\pi}{2L} \,\zeta(-1) = -\,\frac{\pi}{24 L}, \]
 where $\zeta$ is the original Riemann zeta function.
Therefore, a comment is needed upon the relation of
  generalized zeta functions to the approach expounded
here.

 Let $s$ be a complex parameter, and consider the operator function
  $f_s(H) \equiv  H^{-s}$.
 Then  the zeta function for $H$ is def\/ined by
 $\zeta(s,H) \equiv\Tr f_s(H)$
 and extended by analytic continuation to values of~$s$ for which
\[\zeta(s,H) = \sum_{n=1}^\infty \lambda^{-s} \]
 does not converge.
 Note that
 \begin{gather}
\zeta(s,H) = \zeta(2s,\sqrt H). \label{zetas} \end{gather}
On the other hand,
these zeta functions are related to our integral kernels by
\[ \int_0^\infty t^{s-1} T(t,H) \, dt
 = \Gamma(s)\zeta(s,\sqrt H) \]
 and a similar equation with $K(t,H)$.
 It follows from \eqref{heattr} and \eqref{cyltr} that
  $b_n$ and $e_n$ are  residues at  poles of
$\Gamma(s)\zeta(s,H)$
 (at $s=\frac12(d-n)$)
  and  $\Gamma(s)\zeta(s,\sqrt H)$
 (at $s=d-n$),
  respectively.  So when there are no logarithmic terms,
 \eqref{zetas} implies
\[\Gamma\left (\frac{d-n}{2}\right )^{-1}b_n = \frac12\,\Gamma(d-n)^{-1}
e_n. \]
 (This observation is analogous to one by Gilkey \cite{gilkey4th}
concerning higher-order operators.)
Now when $d-n$ is odd and negative, $\Gamma(d-n)$ has a pole where
 $\Gamma\bigl(\frac12(d-n)\bigr)$ does not;
 the information in the corresponding $e_n$ is thereby expunged
 from the heat-kernel expansion, and that is how~\eqref{heattr}
contains less information than \eqref{cyltr}.
In that case, $e_n$ ceases to be a \emph{residue} of the zeta
function and becomes  a \emph{value} of zeta at a regular point~-- a more subtle object to calculate.
(Logarithmic terms give rise to coinciding poles of $\zeta$ and
$\Gamma$.)

 \section{Vacuum energy in a vector bundle from periodic orbits}
   \label{sec:twist}

Let us concentrate now on
 the model studied in \cite{FGR} and summarized above.
 $H =- \od{^2}{x^2}$ acts in $L^2(0,L)$ with the modif\/ied periodic
boundary conditions \eqref{twisbc}, representing a nontrivial
holono\-my in a line bundle over the circle.
 For def\/initeness take $\theta\in [0,2\pi)$.
  (A gauge transformation,
 $\tilde\varphi(x)\equiv e^{i\theta x/L}\varphi (x)$,
 converts this problem to
``Bohm--Aharonov'' form with ordinary periodic boundary conditions
and a nonzero, but pure gauge, vector potential.)
Eigenfunctions must be proportional to $e^{ik_j x}$ with
 $k_j = (2\pi j +\theta)/L$, $j\in \mathbf{Z}$;
  the eigenvalues for positive and negative $k$ then form the two
  sequences
 \eqref{twiseig}.

 Because of the spatial homogeneity of the model,
 there is no distinction, except a factor $L$, between total
energy and local energy density.
 Although vacuum energy is barely mentioned in~\cite{FGR},
\eqref{twisen}
is equivalent to either of the two main results of that paper,
 the cylinder kernel for $H$ and its small-$t$ diagonal expansion
\cite[(9) and (13)]{FGR}, and the f\/irst nonlocal Riesz mean of the
eigenvalue density with respect to frequency $\omega$
 \cite[equation above (30)]{FGR}.
Here I  rederive~\eqref{twisen} in a third way, in the framework of
periodic-orbit theory \cite{BB3,colin,DG}
  (but using the exact formula for the Green
function, not a stationary-phase approximation).

 The Green function (resolvent kernel)
 for the time-independent Schr\"odinger equation
in inf\/inite one-dimensional space is
 \[G_\infty(\omega^2,x,y) = \frac i{2\omega} \, e^{i\omega|x-y|}.
\]
 (By convention, for $\omega^2$ on the positive real axis $G_\omega$ is
def\/ined as the limit from above ($\omega \mapsto \omega+i\epsilon$), or,
equivalently, by the outgoing radiation condition.
Also, let us adopt the convention
 $(H_x-\lambda)G(\lambda,x,y) = +\delta(x-y)$,
 where much of the physical literature has the opposite sign.)
  The Green function for the bundle is constructed as a sum over all
 classical paths connecting $y$ to $x$ in the covering space;
 in this simple one-dimensional problem this construction reduces
to the traditional ``method of images''\negthinspace.
 When $y=x$ the paths become periodic orbits, and there is one for
each $n\in \mathbf{Z}$, with length $|n|L$.
  Since there are no boundaries to ref\/lect the paths, in this model
  there is no distinction between periodic orbits and more general
  closed paths (with dif\/ferent initial and f\/inal velocity),
 in keeping with the previous observation that energy and energy
density are the same thing.
 The Green function $G=G_{L,\theta}$ is
 \[G(\omega^2,x,y)= \sum_{n=-\infty}^\infty G_\infty(\omega^2,x,y+nL)
  e^{in\theta}.\]
 (The sum converges distributionally and has the desired
twisted-periodicity property.)

 The starting point of periodic-orbit theory is that the density
of eigenvalues, as a function of~$\lambda$, is
 \[ \sum_j \delta(\lambda-\lambda_j)  = \frac1{\pi}
 \IIm\Tr G(\lambda). \]
 The trace is an integration over $x$, amounting here to a factor $L$.
  It is more convenient to work either with the density with
respect to $\omega=\sqrt{\lambda}$,
 which carries an additional factor $2\omega$,
 or with the eigenvalue counting function $N$,
 which is the same quantity whether $\lambda$ or
$\omega$ is used as independent variable.
We have
 \begin{align*}
  \IIm G(\omega^2,x,x) &=\IIm
 \sum_{n=-\infty}^\infty  \frac i{2\omega} \,
 e^{i\omega|n|L}  e^{in\theta} \\
&= \frac1{2\omega} +
\frac1{2\omega}\sum_{n=1}^\infty [\cos(\omega nL +n\theta)
 + \cos(\omega nL-n\theta)]
  .\end{align*}
 (One can identify $\omega|n|L$ as the \emph{action} of the $n$th
periodic orbit.
 There is no Maslov index in this problem, but there \emph{is}
 a phase shift $n\theta$ from the nontrivial holonomy.)
 So the integrated eigenvalue density is (for $\omega>0$)
\begin{align} N(\omega) &= \frac L{\pi} \int 2\omega\,d\omega \,
   \IIm  G(\omega^2,x,x) = \frac{L\omega}{\pi} + \frac1{\pi}\sum_{n=1}^\infty \frac1n
 [\sin(\omega nL +n\theta) + \sin(\omega nL-n\theta)]\nonumber  \\
 &\equiv N_\mathrm{av}(\omega) + N_\mathrm{osc}(\omega)
. \label{twisN}\end{align}
 Here $N_\mathrm{av}(\omega)/L = \omega/\pi$
 (which came from the orbit of zero length)
 is the density of states per unit
length that would exist in inf\/inite space, and $N_\mathrm{osc}(\omega)$
 describes the bunching of spectrum caused by the existence of
closed orbits.
        (The contribution of orbit $n$ to \eqref{twisN} includes an
 $\omega$-independent constant of integration,
  $-\sin(n\theta)/n\pi$, which cancels with the corresponding term
  from orbit $-n$.)

 Before turning to the vacuum energy, let's digress to see how
\eqref{twisN}  reproduces the known eigenvalues.
 It is known \cite[1.441.1, 9.627.1]{GR}
  that for $0< z<2\pi$
 \[\sum_{n=1}^\infty \frac{\sin {nz}}{n} = {\pi-\frac z2}
 = -\pi B_1\left (\frac{z}{2\pi}\right ), \]
where $B_1$ is the f\/irst Bernoulli polynomial;
elsewhere, the sum def\/ines a $2\pi$-periodic sawtooth function,
vanishing at the discontinuity points $z=2j\pi$.
 Thus, for $\omega>0$,
 \[N_\mathrm{osc}(\omega)
 = \frac1{2\pi} [f(\pi-\omega L-\theta) +f(\pi-\omega L+\theta)] \]
 where $f(\zeta)$
  is the $2\pi$-periodic extension of the function equal to
 $\zeta$ when $-\pi<\zeta<\pi$ and equal to $0$ at the endpoints.
 Hence $f(\zeta)$  jumps by $-2\pi$ at each odd multiple of $\pi$.
 In more transparent terms, the eigenvalue density is
  (since $N_\mathrm{av}$ and $N_\mathrm{osc}$ vanish for $\omega\le0$)
 \begin{gather*} \rho(\omega) \equiv \od N{\omega}  = \frac L{\pi} + N'_\mathrm{osc}(\omega) = \sum_{j=1}^\infty \delta\left (\omega-\frac{2j\pi-\theta}{L}\right )
 + \sum_{j=0}^\infty \delta\left (\omega-\frac{2j\pi+\theta}{L}\right ),
 \end{gather*}
 in precise agreement with the eigenvalues calculated already by
elementary means.
Of course, in higher-dimensional problems such exact results are
not to be expected.

 The renormalized vacuum energy is the contribution of $N'_\mathrm{osc}$
 to $\int_0^\infty \frac12 \, \omega \rho(\omega) \, d\omega$.
 Returning to \eqref{twisN}, and considering the local energy density, we
have (ignoring analytical technicalities for the moment)
 \begin{gather*} E(x)
 =\int_0^\infty \frac1{2L} \,\omega \,
 N'_\mathrm{osc}(\omega)\, d\omega
 =  \frac1{2\pi}\sum_{n=1}^\infty  \int_0^\infty
 [\cos(\omega nL +n\theta) + \cos(\omega nL-n\theta)]  \,\omega\,d\omega .
 \end{gather*}
 Consider just one term, and integrate to a f\/inite upper limit:
 \[\int_0^\Omega \cos (\omega nL + n\theta)\, \omega\,d\omega
 = \frac{\Omega}{nL}\,\sin(\Omega nL + n\theta)
 + \frac1{(nL)^2}\, \cos (\Omega nL+n\theta)
 -\frac1 {(nL)^2}\, \cos (n\theta). \]
 As $\Omega\to \infty$, the f\/irst two terms oscillate with zero mean.
 If we assume for the moment that they can be ignored, we have
\begin{gather} -\,\frac1 {2\pi(nL)^2}\, \cos (n\theta)
 \label{twisenden}\end{gather}
 as the contribution to the energy density from one of the two
periodic orbits of length $nL$;
 that from the other orbit comes out the same.

 Again from \cite[1.443.3, 9.627.2]{GR},
for $0< \theta<2\pi$
 \[\sum_{n=1}^\infty \frac{\cos {n\theta}}{n^2} =
 \frac{\pi^2}6 - \frac{\pi \theta} 2 + \frac{\theta^2}4
 = \pi^2 B_2\left (\frac\theta{2\pi}\right ), \]
where $B_2$ is the second Bernoulli polynomial.
 The total energy density is thus
 \[ E(x)= - \,\frac\pi{L^2} \, B_2\left (\frac\theta{2\pi}\right )
 = \frac {E_\theta} L, \]
$E_\theta$ given by \eqref{twisen}, as was to be verif\/ied.

 It is noteworthy that the vacuum energy \eqref{twisenden}
  associated with a
single spectral oscillation depends critically, in sign as well as
magnitude, on the phase $n\theta$ of the oscillation.
 Algebraically, \eqref{twisenden} comes entirely from the lower
limit of the integration over $\omega$, and an ef\/fective lower
cutof\/f on that integration would appear to change the phase.
 This is a matter of great concern for the extendability of the
theory, since in more general circumstances the Gutzwiller spectral
oscillations arise from stationary-phase approximations that are
not justif\/ied at low frequency.
 It has often been observed, however, that periodic-orbit
calculations reproduce low-lying eigenvalues more accurately than
they have any right to do.

 The problem at high frequency, in contrast, is more apparent than
real.
 The integral over~$\omega$ can be def\/ined by Riesz--Ces\`aro
summation of order 2:
 A lengthy exercise in integration by parts shows that
 \[ \int_0^\Omega \left (1-\frac{\omega}{\Omega}\right )^2
 \cos (\omega nL+n\theta)\,\omega\,d\omega =
-\,\frac1 {(nL)^2}\, \cos (n\theta) +O(\Omega^{-1}). \]
Taking the limit $\Omega\to\infty$ now yields~\eqref{twisenden}.
 Clearly, the interchange of integration and summation in the
calculation of $E(x)$ is now also justif\/iable.

 Another approach leading to the same conclusion is Abel summation,
which amounts to f\/inding the contribution of each spectral
oscillation to the cylinder kernel.
 We are after the small-$t$ behavior of
 \[\int_0^\infty \frac12 \, \omega \rho(\omega) e^{-\omega t} \,d\omega
 = -\,\frac12 \,\od {}t \int_0^\infty \rho(\omega) e^{-\omega t}\,
d\omega. \]
 Using the integral
 \begin{gather}
 \int_0^\infty \cos (a\omega-b) e^{-\omega t}\, d\omega =
 \frac{t}{t^2+a^2}\, \cos b + \frac{a}{t^2+a^2}\, \sin b,
\label{coslap} \end{gather}
 one f\/inds that the contribution of the two orbits with length $nL$
 to $\int_0^\infty \rho(\omega) e^{-\omega t}\, d\omega$ is
 \[ \frac{2L} \pi \,\frac{t}{t^2+ (nL)^2} \,\cos (n\theta). \]
 (The two contributions are not equal; their sine terms cancel.
But because the sine term in \eqref{coslap} has no term of order
exactly $t^{+1}$ in its small-$t$ expansion, it would make no
contribution to \eqref{twisenden} even before
  the pairwise cancellation.)
 The sum over $n$ is now absolutely convergent, and at $t=0$
 the expected formulas \eqref{twisenden} and \eqref{twisen} emerge.
 Alternatively, the series at f\/inite $t$ can be summed in closed
form \cite[1.445.2]{GR}, and with the help of some
identities for  hyperbolic functions the result is shown equal to
the diagonal value of the cylinder kernel as
  found previously \cite[(12)]{FGR}.

 If $\theta$ is written $|\theta|$, \eqref{twisen} is valid over
the interval
  $-2\pi\le \theta\le2\pi$.
 Overall, $E_\theta$ is a continuous $2\pi$-periodic function with
cusps at the integer multiples of $2\pi$;
 it is symmetric under ref\/lection about any integer multiple of
$\pi$.
 We have already commented in Section~\ref{sec:general} about the
signif\/icance of its extreme values and its zero.

 \section{Boundary  vacuum energy from closed and periodic
orbits}  \label{sec:bdry}

 We consider a f\/inite interval with either a Dirichlet or a Neumann
boundary condition at each end. Thus
 $H =- \od{^2}{x^2}$ acts in $L^2(0,L)$ on the domain def\/ined by
 \[ u^{(1-l)}(0) =0,\qquad u^{(1-r)}(L) =0, \qquad\text{where }
  l, r \in \{0,1\}. \]
 For $l=1$, the (unnormalized) eigenfunctions are $\,\sin(\omega_j x)$
 with  \[
  \omega_j = \frac{\pi j}L,\quad j\in \mathbf{Z}^+, \quad\text{if $r=1$};
 \qquad
 \omega_j = \frac\pi{L}\left (j+\frac12\right ),\quad j\in \mathbf{N},
 \quad\text{if $r=0$}. \]
 For $l=0$, they are $\,\cos(\omega_j x)$ with $j\in\mathbf{N}$ and
 \[ \omega_j = \begin{cases}
 \displaystyle\frac\pi{L}\left (j+\frac12\right )
  &\text{if $r=1$,} \\
 \noalign{\smallskip}
\displaystyle \frac{\pi j}L & \text{if $r=0$.}
  \end{cases} \]

 The Green function can be constructed from $G_\infty$  by the method
of images, or, equivalently, as a sum over ref\/lected paths as in
\cite[Fig.~1]{SS}.
(I  suppress the arguments $\omega^2$ and $x$ after their f\/irst
appearance.)
\begin{align*} G(\omega^2,x,y) &=
 G_\infty(y) + (-1)^l G_\infty(-y)
 + (-1)^r G_\infty(2L-y) +  (-1)^{l+r}G_\infty(2L+y) \\
 \noalign{\smallskip}
 &\quad{}+ (-1)^{l+r} G_\infty(-2L+y) +(-1)^{2l+r} G_\infty(-2L-y) \\
&\quad
 {} + (-1)^{l+2r} G_\infty(4L-y) + (-1)^{2l+2r} G_\infty(4L+y)
+\cdots \\
 &= \sum_{n=0}^\infty (-1)^{n(l+r)} G_\infty(-2nL+y)
 +\sum_{n=0}^\infty (-1)^{l+n(l+r)} G_\infty(-2nL-y) \\
&\quad{}+ \sum_{n=1}^\infty (-1)^{-l+n(l+r)} G_\infty(2nL-y)
 +\sum_{n=1}^\infty (-1)^{n(l+r)} G_\infty(2nL+y)  \\
 &= \frac{i}{2\omega} \sum_{n=0}^\infty (-1)^{n(l+r)}
 e^{i\omega|2nL+x-y|}
 + \frac{i}{2\omega} \sum_{n=0}^\infty (-1)^{l+n(l+r)}
 e^{i\omega(2nL+x+y)} \\
&\quad{} +\frac{i}{2\omega} \sum_{n=1}^\infty
 (-1)^{-l+n(l+r)} e^{i\omega(2nL-x-y)}
 + \frac{i}{2\omega} \sum_{n=1}^\infty (-1)^{n(l+r)}
  e^{i\omega(2nL-x+y)}
 \end{align*}
 (cf.\ \cite[(2.16)]{SS}).
 When $y=x$, the f\/irst and fourth sums correspond to periodic
paths
 (including the direct path of zero length),
  but the second and
third sums come from paths that are closed but not periodic (i.e.,
the ``particle'' has hit the boundary an odd number of times and
returned to its starting point with reversed velocity).
The $n=0$ term in the f\/irst sum is the direct path;
$n=0$ in the second term and $n=1$ in the third (which will become
$n=-1$ in the next step) are the ``short'' paths that bounce of\/f
the boundary only once.

 We have (always with the understanding that $\omega>0$)
 \begin{align*} \pi\sigma &\equiv 2\omega \IIm G(\omega^2,x,x)
 \\ &=
  \sum_{n=0}^\infty (-1)^{n(l+r)} \cos (2\omega nL)
 + \sum_{n=0}^\infty (-1)^{l+n(l+r)} \cos(2\omega (nL+x))
 \\ &\quad {}
 +\sum_{n=1}^\infty (-1)^{-l+n(l+r)} \cos(2\omega (nL-  x))
 +\sum_{n=1}^\infty (-1)^{n(l+r)} \cos (2\omega nL)  \\
&=  1 + 2\sum_{n=1}^\infty (-1)^{n(l+r)} \cos(2\omega nL)
 +\sum_{n=-\infty}^\infty (-1)^{l+n(l+r)} \cos(2\omega (x+nL)) \\
&\equiv \pi(\sigma_\text{av} + \sigma_\text{per} +\sigma_\text{bdry})
  \equiv \pi(\sigma_\text{av} + \sigma_\text{osc}) .
 \end{align*}
 Because the series are not absolutely
 convergent, we must be wary of cancelling contributions from
dif\/ferent values of $n$.
  For the most part, we want to consider to the latest possible
moment the contribution from each orbit separately, and the
summation is just a reminder of which values of $n$ occur.
 We do pair $n$ with $-n$ or $-(n+1)$ when their
contributions are manifestly identical or parallel.
 We have just done so for the f\/irst and fourth terms in $\sigma$,
which became $\sigma_\text{av}+\sigma_\text{per}$.
 In the second and third terms, $\sigma_\text{bdry}$,
 it will be  natural to pair $n$ $({}>0)$
 with $-(n+1)$,
 since that combines paths that have the same number of
ref\/lections, $2n+1$ (again consult \cite[Fig.~1]{SS}, or just look at
the exponents of $(-1)$ in the formula).
 At f\/ixed $x$, orbits $n$ and $-(n+1)$ do not have the same length,
 but when integrated over $x$ those two sets of orbits collectively
occupy the same part of the length spectrum.

 Let us compare with the case of a single boundary
 (Hilbert space $L^2(0,\infty)$, boundary condition $u^{(1-l)}(0)=0$).
 In that case only the direct path and the short path of\/f the
boundary exist.
 Thus $\sigma_\text{per}=0$, $\sigma_\text{av}$ is unchanged, and
$\sigma_\text{bdry}= \frac{(-1)^l}{\pi}\cos(2\omega x)$ has only one term.

 There are now two routes to follow: We can work with the local
spectral density $\sigma$ (integrating with respect to $\omega $,
summing over $n$, calculating the local energy density, etc.),
 or we can integrate over $x$ f\/irst to get the global eigenvalue
density.  (Unfortunately, ``density'' is ambiguous in this
context, as previously remarked.)

Let's examine the global situation f\/irst.
 The eigenvalue density is
 \[\rho(\omega ) = \int_0^L \sigma(\omega ,x)\, dx =
 \rho_\text{Weyl} +\rho_\text{per} +\rho_\text{bdry}\,, \]
 where
\begin{gather*}
\rho_\text{Weyl} = \int_0^L \sigma_\text{av}\, dx = \frac
L{\pi}, \qquad
 \rho_\text{per} = \frac{2L}\pi \sum_{n=1}^\infty
 (-1)^{n(l+r)} \cos (2\omega nL), \\
\rho_\text{bdry} = \frac{(-1)^l}{2\pi} \sum_{n=-\infty}^\infty
 \frac{(-1)^{n(l+r)}}{\omega } \,
 [\sin(2\omega L(n+1)) - \sin (2\omega Ln)].
 \end{gather*}
 In keeping with previous remarks, we refuse for the moment to take
advantage of the possibility of ``telescoping'' $\rho_\text{bdry}$
 when $l+r$ is even, but we can combine the positive and negative
parts of the series now.
 In the summand
 $\frac{(-1)^{n(l+r)}}{\omega } \,
 [\sin(2\omega L(n+1)) - \sin (2\omega Ln)]$,
 replace $n$ by $-(n+1)$:
 \begin{gather*}
\frac {(-1)^{-(n+1)(l+r)}}{\omega } \,
 [\sin(2\omega L(-n)) - \sin (-2\omega L(n+1))]
 \\
\qquad{} = (-1)^{l+r} \frac{(-1)^{n(l+r)}}{\omega } \,
 [\sin(2\omega L(n+1)) - \sin (2\omega Ln)] ,
 \end{gather*}
 which is the original summand except for the initial sign.
 Therefore, if $l+r$ is odd (the two boundary conditions are
dif\/ferent), $\rho_\text{bdry}=0$; but if $l+r$ is even,
 \[ \rho_\text{bdry} =
\frac{(-1)^l}{\pi} \sum_{n=0}^\infty \frac1\omega \,
[\sin(2\omega L(n+1)) - \sin (2\omega Ln)]. \]
(The local counterpart of this observation is that
 the simultaneous transformation $n\mapsto -(n+1)$,
 $x\mapsto L-x$ leaves the summand invariant up to the sign.
Together with the sign ${(-1)^l}$, this indicates that
  the ef\/fects that are localized near the boundaries
 are  equal and opposite for Dirichlet and Neumann boundaries.)

 The corresponding equation for the half-line with a single
boundary is formally
\begin{gather}
 \rho_\text{bdry} =\frac{(-1)^l}{\pi}\int_0^\infty \cos(2\omega x)\,dx
 = \frac{(-1)^l}{4} \delta(\omega ) .
 \label{rhobdry}\end{gather}
 Of course $\rho_\text{av}$ is inf\/inite in that case because of the
inf\/inite volume and continuous spectrum, but $\rho_\text{bdry}\,$,
 being associated with the boundary, is a localized, f\/inite
contribution that can be studied separately.

\smallskip

\noindent
 {\bf Remark.} The distributional integral \eqref{rhobdry} has
been thoroughly studied in \cite{euler}.  The second equality in
\eqref{rhobdry} is correct under the convention that
 \begin{gather}
 f(0)\equiv \int_0^\infty \delta(\omega) f(\omega)\, d\omega
 = \int_{-\infty}^\infty \delta(\omega)H(\omega) f(\omega)\, d\omega,
 \label{deldef}\end{gather}
where $H$ is the unit step function.
 An alternative convention is that
a delta function at an endpoint
of an interval of integration yields only half the value of the
test function at that point;
 otherwise put, the   integration is extended over all
 $\omega \in \mathbf{R}$
and the integrand is interpreted according to the rule
$\delta(\omega )\theta(\omega ) = {\textstyle \frac12} \delta(\omega )$.
The choice is somewhat arbitrary \cite{euler}, but \eqref{deldef} has
the advantage that an eigenvalue $\omega_0{}\!^2$ always
corresponds in $\rho$ to a unit-normalized delta function,
 $\delta(\omega-\omega_0)$,  even when $\omega_0=0$.

 In a more general problem, the integrals over $\omega $ would need to be
evaluated in a stationary-phase approximation.
 It would then be argued that only periodic orbits contribute, so
$\rho_\text{bdry}$ would be set to zero.
A formal justif\/ication for this approximation is that the neglected
terms are of higher order in Planck's constant when nonrelativistic
quantum-mechanical units are used in $H$.
 In  relativistic vacuum-energy calculations in natural units, the
manifestation of this observation is that $\rho_\text{bdry}$
 is suppressed relative to $\rho_\text{av} +\rho_\text{per}$ by a
factor $1/\omega L$, which is small in the high-frequency regime where
WKB-type asymptotics would be valid.
 A less drastic approximation is to keep only the two short orbits:
\[\rho_\text{bdry} \approx \frac{(-1)^l}{\pi}\,
 \frac{\sin(2\omega L)}{\omega }.
\]
 This is plausible because the boundary ef\/fect should come only
from points close to the boundary (cf.\ the local calculations
below and the single-boundary equation above).

 The eigenvalue counting function $N(\omega )$ is zero for
 $\omega <0$ and  $\int_0^\omega  \rho$ for $\omega >0$.
 Therefore (for $\omega>0$),
 \begin{gather*} N_\text{Weyl} = \frac{L\omega }{\pi}, \qquad
 N_\text{per} = \frac1{\pi} \sum_{n=1}^\infty \frac{(-1)^{n(l+r)}}{n}
 \,\sin(2\omega nL), \\
 N_\text{bdry} = \begin{cases}
\displaystyle \frac{(-1)^l}{\pi} \sum_{n=0}^\infty \int_0^\omega
 \frac{\sin(2\omega L(n+1)) -\sin(2\omega Ln)}{\omega }\, d\omega
&\text{if $l+r$ is even,}
 \\ 0 &\text{if $l+r$ is odd.}
\end{cases}
\end{gather*}
 We contemplate each of these in turn.

 $N_\text{Weyl}$ is exactly as expected.

 If $l+r$ is odd, by \cite[1.441.3]{GR} we have
\[ N_\text{per} = \frac1{\pi} \sum_{n=1}^\infty \frac{(-1)^n}{n}
 \,\sin(2\omega nL)
=-\, \frac1{2\pi}\, f(2\omega L) \]
with the same $f(\zeta)$ as def\/ined in Section~\ref{sec:twist}.
 That is,  if we temporarily forget the restriction to $\omega >0$,
 $N_\text{per}$ equals $- \frac{2\omega L}{2\pi} = - N_\text{Weyl}$
 for $\omega \in \bigl(-\frac{\pi}{2L}, \frac{\pi}{2L}\bigr)$,
 vanishes at the endpoints, and is periodic thereafter;
 this function jumps upward by $1$ at each (positive) odd multiple
of  $\frac\pi{2L}$, which we know to be the correct eigenvalues for
these problems.
 (Another way of saying this sort of thing is that $N_\text{per}$ is
equal to the negative of $N_\text{Weyl}\,$,
 plus a series of unit step functions located at the eigenvalues.)
 At $\omega =0$ this function equals $0$, so the complete
$N_\text{per}$ is continuous there.
 For consistency, $N_\text{bdry}$ must turn out to be $0$ when $l+r$
is odd (and it does).

 If $l+r$ is even,
\[N_\text{per} = \frac1{\pi} \sum_{n=1}^\infty  \frac{\sin(2\omega nL)}{n}
= \frac1{2\pi}\, f(\pi-2\omega L), \]
 so the jumps occur at the (positive) integer multiples of
$\frac{\pi}{L}$, as they should.
 The only complication is at $\omega =0$, where this function approaches
$\frac12$ from the right.
 Since the complete $N_\text{per}$ is $0$ for $\omega <0$,
 there is only \emph{half} a step function at $0$.
 In the full $N$, there should be a complete unit jump in the
Neumann problem ($l=0$) and no jump at all in the Dirichlet problem
 ($l=1$).
 These corrections must come, of course, from the boundary term.

 For it we have
  \begin{align*} N_\text{bdry}
 &=\frac{(-1)^l}{\pi} \sum_{n=0}^\infty
 \left [\int_0^{2L\omega (n+1)} \frac{\sin u}{u}\,du -
 \int_0^{2L\omega n} \frac{\sin u}{u}\,du\right ] \\
 &= \frac{(-1)^l}{\pi} \sum_{n=0}^\infty
\int_{2L\omega n}^{2L\omega (n+1)}\frac{\sin u}{u}\,du
 = \frac{(-1)^l}{\pi} \int_0^\infty \frac{\sin u}{u}\,du
 = \frac{(-1)^l}{2}
 \end{align*}
 (see, e.g., \cite[3.721.1]{GR}).
This holds for $\omega >0$; thus $N_\text{bdry}$ heals the
discontinuity at $\omega =0$ in~$N_\text{per}$ when $l=1$ and
strengthens it to a unit jump when $l=0$, as expected.

 Also noteworthy, although not unexpected,
  are that $N_\text{bdry}$ is independent of~$L$~-- it is associated with the physics of the boundary, not the
f\/initeness of the region~--
 and that it is nonoscillatory as a function of $\omega $.
 Thus for $N$
 (and other global quantities) it seems proper to write
\begin{gather*}
N_\text{av} \equiv N_\text{Weyl} + N_\text{bdry},
 \qquad N_\text{osc} \equiv N_\text{per},
 \end{gather*}
 in contrast to the def\/inition of
  $\sigma_\text{av}$ and $\sigma_\text{osc}$.

 If we kept only the
contribution from the short orbits, we would get
 \[N_\text{bdry} \approx \frac{(-1)^l}{\pi}\,\int_0^{2L\omega }
 \frac{\sin u}{u}\,du, \]
 a fair approximation to the correct step function
when $L\omega $ is large.
For comparison,
 our formula for $\rho_\text{bdry}$ on the half-line
  also yields  a step function at the
origin of half that magnitude, $\frac{(-1)^l}{4}\,H(\omega )$,
as  one should expect \cite{BG} for only a single boundary.

 Now consider the regularized vacuum energy
\[E(t) = -\, \od{}t\, \frac12 \int_0^\infty \rho(\omega )
 e^{-\omega t}\, d\omega
 \equiv E_\text{Weyl} + E_\text{per}+E_\text{bdry},\]
 where
 \[E_\text{Weyl}(t) = -\, \frac L{2\pi}\, \od{}t \,\frac1t
 = \frac L{2\pi t^2}, \]
 and
\begin{align*}
  E_\text{per}(t) &= -\, \frac L{\pi}\, \od{}t \,
 \sum_{n=1}^\infty (-1)^{n(l+r)} \int_0^\infty \cos(2\omega nL)
 e^{-\omega t}\, d\omega  \\
 &= -\, \frac L{\pi}\, \od{}t \,
 \sum_{n=1}^\infty (-1)^{n(l+r)} \, \frac t{t^2+(2nL)^2}
= -\, \frac1{2\pi} \,\od{}t \,
 \sum_{n=1}^\infty (-1)^{n(l+r)} \, \frac {t/2L}{(t/2L)^2+n^2}.
 \end{align*}
  These series can be evaluated by
 \cite[1.217.1,2]{GR} or their generalizations
\cite[1.445.2,3]{GR}:
 In the even case,
 \[ E_\text{per}(t) =   -\, \frac1{2\pi} \,\od{}t \,
 \left [\frac{\pi}2 \, \coth \biggl(\frac{\pi t}{2L}\biggr)
 - \frac {L}{t} \right ]
=\frac{\pi}{8L} \,\csch^2\biggl(\frac{\pi t}{2L}\biggr)
 -\frac L{2\pi t^2}.
 \]
 In the odd case,
\[ E_\text{per}(t) =   -\, \frac1{2\pi} \,\od{}t \,
 \left [\frac{\pi}2 \, \csch \biggl(\frac{\pi t}{2L}\biggr)
 - \frac {L}{t} \right ]
 =\frac{\pi}{8L} \,\csch\biggl(\frac{\pi t}{2L}\biggr)
\coth\biggl(\frac{\pi t}{2L}\biggr)
 -\frac L{2\pi t^2}.
 \]
Expand these in  power series in $t$, using
\cite[1.411.8,12]{GR}:
\[ \csch z = \frac1z -\frac z6 + O(z^3),\qquad
 \coth z = \frac1z + \frac z3 + O(z^3). \]
 The terms of order $t^{-2}$ cancel, and we get
 \[ E_\text{per}(t) = \begin{cases}
\displaystyle -\frac{\pi}{24L} + O(t^2) &\text{if $l+r$ is even,}\\
 \noalign{\smallskip}
 \displaystyle\hphantom{-} \frac{\pi}{48L} + O(t^2)
 &\text{if $l+r$ is odd.}
 \end{cases}\]
 The f\/irst of these gives the well known renormalized vacuum energy
\eqref{inten} for the one-dimensional Dirichlet problem ($l=r=1$).
 It is also correct for the Neumann case, when any energy associated
with the indiscretely quantized zero mode is neglected.
 The extra factor of $-\frac12$ in the mixed case is just like that
for the antiperiodic case in Section~\ref{sec:twist} and \cite{isham}
 and similar to the factor
of $-\frac78$ in the mixed case
in three-dimensional electromagnetism \cite{boyer};
  this  family of formulas can be  obtained at the eigenfunction
  level
by doubling the interval and removing the contribution of the even
modes~\cite{dowker}.

It remains to investigate the boundary energy.
 As usual it will be zero in the odd case.
 For the even case we have
 \begin{align*} E_\text{bdry}(t) &= -\,\frac{(-1)^l}{2\pi} \,\od{}t
 \sum_{n=0}^\infty \int_0^\infty
 [\sin(2\omega L(n+1)) - \sin (2\omega Ln)]\,
 \frac{e^{-\omega t}}{\omega }\,d\omega  \\
&= \frac {(-1)^l}{2\pi}  \sum_{n=0}^\infty
 \int_0^\infty [\sin(2\omega L(n+1)) - \sin (2\omega Ln)]
 e^{-\omega t}\,d\omega  \\
 &= \frac{(-1)^l}{2\pi}    \sum_{n=0}^\infty
\left [ \frac{2L(n+1)}{t^2 + (2L(n+1))^2} - \frac{2Ln}{t^2 + (2Ln)^2}
\right ].
 \end{align*}
 This series is conditionally convergent  and telescopes to zero.
Alternatively,  put the two terms in the summand over a common
denominator:
 \[ E_\text{bdry}(t) = \frac{(-1)^l}{2\pi}  \sum_{n=0}^\infty
 \frac{2Lt^2 -(2L)^3 n(n+1)}
 { [t^2 + (2L(n+1))^2][t^2 + (2Ln)^2]} . \]
 The term for $n=0$ is
\[\frac{(-1)^l}{\pi} \,\frac L{t^2+4L^2} =
 \frac{(-1)^l}{4\pi L} +O(t^2). \]
In the other terms it is legitimate to expand the denominators
before summing:
\begin{gather*}
  \frac{(-1)^l}{2\pi}  \sum_{n=1}^\infty
\frac {2Lt^2 -(2L)^3 n(n+1)}{ (2L)^4 n^2(n+1)^2}
 \left [1 - \frac{t^2}{ (2Ln)^2} +\cdots\right ]
 \left [1 - \frac{t^2}{ (2L(n+1))^2} +\cdots\right ]
 \\
\qquad{} = -\,\frac{(-1)^l}{4\pi L} \, \sum_{n=1}^\infty
      \frac1{n(n+1)}+O(t^2) = -\,\frac{(-1)^l}{4\pi L}+O(t^2),
 \end{gather*}
  the numerical sum being a textbook example
 \cite[pp.\ 612--613]{stewart}
 of a telescoping
series that converges to $1$.
  So all these terms exactly cancel the
$n=0$ term in the limit of small $t$, and $ E_\text{bdry}(0)=0$.
This result was not entirely obvious, since one might expect
boundary energies (possibly inf\/inite) at both ends with the same
sign.

 In the approximation of keeping only the short orbits, one gets a
nonzero (and $L$-dependent) result,
  $ E_\text{bdry}(0)= \frac{(-1)^l}{4\pi L}\,$.
 For the half-line we have
 \begin{align*} E_\text{bdry}(t) &=
 -\frac12 \,\od{}t \int_0^\infty \frac{(-1)^l}{4}\,\delta(\omega )
 e^{-\omega t}\, d\omega
\quad\text{or}\quad
 \frac12 \int_0^\infty \frac{(-1)^l}{4}\,\delta(\omega )\,
 \omega e^{-\omega t}\, d\omega \\
 &= 0. \end{align*}

 Now we turn to local quantities.
 First, integrate $\sigma$ to get a local analogue of the counting
function.
 (This is the inverse Laplace transform of the diagonal value of
the heat kernel; it is the quantity called $\mu^{00}$ in
\cite{lgacee1}.)
 As expected,
 \begin{gather*}
 \int \sigma_\text{av}\, d\omega  = \frac \omega {\pi},\qquad
\int \sigma_\text{per}\, d\omega  = \frac 1{\pi L}
 \sum_{n=1}^\infty \frac{(-1)^{n(l+r)}}{n}\, \sin(2nL\omega )
 = \frac1L\, N_\text{per}.
 \end{gather*}

 The boundary term is
\[\int \sigma_\text{bdry}\, d\omega  = \frac 1{\pi }
 \sum_{n=-\infty}^\infty (-1)^{l+n(l+r)} \,
\frac {\sin(2\omega (x+nL)}{2(x+nL)} . \]
 Integration of it over $x$ yields the same $N_\text{bdry}$ found
before.
 The analogous calculations for the half-line give
\[\int \sigma_\text{bdry}\, d\omega  = \frac{(-1)^l}{2\pi}\,
\frac {\sin(2\omega x)}{ x},\]
 which is precisely the $n=0$ term in the sum above, and
\[ N_\text{bdry} = \frac{(-1)^l}{2\pi}\,  \int_0^\infty
\frac {\sin(2\omega x)}{x}\,dx = \frac{(-1)^l}{4}, \]
in agreement with our previous result \eqref{rhobdry} for this case.

At this point let's pause to compare the results with those from a
direct summation of the eigenfunction expansion, specializing to
the doubly Dirichlet case, $l=r=1$.
 The diagonal value of the (integrated) spectral kernel is
\[\sum_{(\pi j/L)\le \omega } |\varphi_j(x)|^2
 =\sum_{j=1}^\floor{\omega L/\pi} \frac2L\, \sin^2\biggl(\frac{\pi jx}{L}
\biggr), \]
 which reduces after some calculation to
 \[\frac1{2L} + \frac1L\,\floor{\omega L/\pi}
 -\frac1{2L} \sum_{j=-\floor{\omega L/\pi}}^\floor{\omega L/\pi}
 e^{2\pi ijx/L}. \]
 The f\/irst two terms are precisely the step function
 $\frac1L(N_\text{av}+N_\text{per})$.
  The sum in the f\/inal term is the Dirichlet kernel
  introduced in any  rigorous textbook on Fourier series
 (e.g., \cite[p.~22]{helson}).
 Thus
\[ \int \sigma_\text{bdry}\,d\omega  = -\,\frac1{2L}\,
\frac{\sin \left ( (1+2\floor{\frac{\omega L}{\pi}})\frac{\pi x}{L}\right)}
 { \sin\left (\frac{\pi x}{L}\right )}. \]
 When $\omega $ is large, this function develops sharp peaks near $x=0$
and $x=L$, in keeping with the boundary-ef\/fect picture we have had
in  mind all along.
 But instead of performing this sum, we can relate it to the
closed-orbit calculation by the Poisson summation formula:
\begin{align*} -2L \int \sigma_\text{bdry}\,d\omega  &=
\sum_{j=-\floor{\omega L/\pi}}^\floor{\omega L/\pi}  e^{2\pi ijx/L}
= \sum_{j=-\infty}^\infty e^{2\pi ijx/L} \theta(\omega L-\pi|j|) \\
 &= \sum_{n=-\infty}^\infty \int_{-\infty}^\infty dj\,
 e^{2\pi i jn} e^{2\pi ijx/L} \theta(\omega L-\pi|j|) \\
 &= \sum_{n=-\infty}^\infty
 \int_{-\floor{\omega L/\pi}}^\floor{\omega L/\pi}
e^{2\pi ij (n+x/L)}\, dj = \sum_{n=-\infty}^\infty \frac{\sin(2\omega (x+nL))}{x+nL}.
 \end{align*}

Finally, we calculate the local energy density.
 The  def\/inition of  the energy density  of the massless scalar
f\/ield in spatial dimension~1 in f\/lat space is
\[T_{00}(x) =  \frac12\left [
 \left (\pd \phi t\right )^2 + \left (\pd \phi x\right )^2
 -4\xi \left [\left (\pd \phi x\right )^2
 + \phi\, \pd{^2\phi}{x^2} \right ]\right ], \]
 where $\xi$ is called the \emph{conformal coupling parameter};
 dif\/ferent values of $\xi$ correspond to dif\/ferent theories of the
coupling of the f\/ield to gravity, but in f\/lat space they are
physically and mathematically equivalent apart from this one
def\/inition.
 The term multiplied by $-4\xi$ equals
\[
\pd{}x \left (\phi\, \pd \phi x\right ),
\]
 which classically vanishes upon integration by parts in free space or
 under either
Dirichlet or pure Neumann boundary conditions.
 In dimension~1 there are only two distinguished values of~$\xi$,
 namely $0$ (the conformal, or Yamabe, choice) and $\frac14$
 (which is needed for energy conservation in the presence of
general boundary conditions if singular surface terms are to be avoided
 \cite{lebedev1,RS,systemat}).
 The energy density for general $\xi$ is a convex combination of
these two special cases.
 In the conformal case in dimension~1
 the contribution of each normal mode to the
vacuum expectation value is independent of $x$, so we know
that $T_{00}$ is just $E/L$.
 (In particular, there is no boundary contribution.)
 In the case $\xi=\frac14$ the contribution of the space
derivatives is identical to that of the time derivatives, so we can
write
\[ T_{00}(t,x)\equiv E(t,x) =
 -\frac12 \, \pd{}t \int_0^\infty \sigma(\omega ) e^{-
\omega t}\,d\omega
 \equiv E_\text{Weyl}(t)+E_\text{per}(t)+E_\text{bdry}(t,x),
 \]
 which now is indeed the energy formula with the integration stripped of\/f.

Clearly, for the f\/irst two terms we get the same old result,
\[E_\text{Weyl}(0)+E_\text{per}(0) = \frac EL\,.\]
 The interesting term is
\begin{align*} E_\text{bdry}(t,x) &=
-\, \frac{(-1)^l}{2\pi} \,\pd{}t \sum_{n=-\infty}^\infty
 (-1)^{n(l+r)} \int_0^\infty \cos (2\omega (x+nL))e^{-\omega t}\,
  d\omega  \\
 &= -\,\frac{(-1)^l}{2\pi} \,\pd{}t \sum_{n=-\infty}^\infty
 (-1)^{n(l+r)}  \frac t{t^2+4(x+nL)^2} \\
 &=\frac{(-1)^l}{2\pi}  \sum_{n=-\infty}^\infty (-1)^{n(l+r)}\,
\frac {t^2 - 4(x+nL)^2 }{[t^2 + 4(x+nL)^2]^2}\,.
 \end{align*}
 These sums do not appear in \cite{GR},  but
 {\sl Mathematica\/} \cite{mma} evaluates them in terms of hyperbolic
functions of complex argument.
 Some hindsight reveals what is going on:
 The summations we used to evaluate the total energy (ef\/fectively
the present sums with $x=0$) can be written
\[ \frac{\pi}a \,\coth(\pi a) = \sum_{n=-\infty}^\infty
\frac1{n^2+a^2}, \qquad
   \frac{\pi}a \,\csch(\pi a) = \sum_{n=-\infty}^\infty
\frac{(-1)^n}{n^2+a^2}. \]
Factoring the terms on the right, we see that these are just
 Mittag--Lef\/f\/ler expansions of the hyperbolic functions in terms of
simple poles:
 \[ 2\pi \coth(\pi a)= \sum_{n=-\infty}\left ( \frac1{a+in}
 +\frac1{a-in}\right )  \]
and a similar formula for $\,\csch$.
 So the summands with a general quadratic in the denominator can be
treated by displacing the argument of the hyperbolic functions and
letting $n\mapsto -n$ in half the terms:
\begin{align*} \sum_{n=-\infty}^\infty \frac1{(n+b)^2+a^2}& =
 \frac{\pi}{2a} [\coth(\pi(a+ib))+\coth(\pi(a-ib))], \\
    \sum_{n=-\infty}^\infty \frac{(-1)^n}{(n+b)^2+a^2}& =
 \frac{\pi}{2a} [\csch(\pi(a+ib))+\csch(\pi(a-ib))].
 \end{align*}
 (These formulas are actually improvements on {\sl Mathematica\/}'s
 output, though equivalent.)

Thus when $l+r$ is even,
 \[ E_\text{bdry}(t,x) = -\,\frac{(-1)^l}{8L} \pd{}t
 \left [\coth \left( \frac{\pi t}{2L} + \frac{i\pi x}{L}\right )
 + \coth \left (\frac{\pi t}{2L} - \frac{i\pi x}{L}\right )\right ],
\]
 and when $l+r$ is odd, the formula is the same with $\,\coth\,$
 replaced by $\,\csch$.
Dif\/ferentiating f\/irst, and then working out the complex arithmetic,
one gets in the even case
 \[ E_\text{bdry}(t,x) =\frac{(-1)^l\pi}{8L^2} \,\frac
 {\sinh^2\left (\frac{\pi t}{2L}\right )
 \cos^2\left (\frac{\pi x}{L}\right )
 -\cosh^2\left (\frac{\pi t}{2L}\right )
 \sin^2\left (\frac{\pi x}{L}\right )
 }{
 \left [\sinh^2\left (\frac{\pi t}{2L}\right ) \cos^2
 \left (\frac{\pi x}{L}\right )
 +\cosh^2\left (\frac{\pi t}{2L}\right ) \sin^2
 \left (\frac{\pi x}{L}\right )\right ]^2 }
 \]
 and in the odd case
 \[ E_\text{bdry}(t,x) = \frac{(-1)^l\pi}{8L^2} \, \frac
 {\cosh\left (\frac{\pi t}{2L}\right )\cos\left (\frac{\pi x}{L}\right )
 \left [\sinh^2\left (\frac{\pi t}{2L}\right )
 -\sin^2\left (\frac{\pi x}{L}\right )
 + 2 \sinh^2\left (\frac{\pi t}{2L}\right ) \sin^2
 \left (\frac{\pi x}{L}\right )\right ]
 }{
 \left [\sinh^2\left (\frac{\pi t}{2L}\right ) \cos^2
 \left (\frac{\pi x}{L}\right )
 +\cosh^2\left (\frac{\pi t}{2L}\right ) \sin^2
 \left (\frac{\pi x}{L}\right )\right ]^2 }
\,. \]

 Specializing to  $l=1$, let us examine the leading terms at
small $t$ and small $x$.
 For the even case,
 \[ E_\text{bdry}(0,x) = \frac{\pi}{8L^2} \,
 \csc^2\left (\frac{\pi x}{L}\right ). \]
 This is the renormalized boundary energy density in~\eqref{intenden}
 (Fig.~\ref{fig:int}).
 Its integral over the whole interval is not even f\/inite, much less
zero as formally expected.
 Near $x=0$ we have
\[ E_\text{bdry}(0,x) = \frac1{8\pi x^2} + \frac{\pi}{24L^2} +
 O(x^2) \]
 (and a corresponding expansion near $x=L$).
 On the other hand, if we expand in $x$ f\/irst we get
\[ E_\text{bdry}(t,x) = -\, \frac{\pi}{8L^2}
 \, \csch^2\left (\frac{\pi t}{2L}\right ) +O(x^2), \]
 the remainder term being nonuniform in $t$.
 At $x=0$ and small $t$ this becomes
\[ E_\text{bdry}(t,0) \approx -\,\frac1{2\pi t^2} \]
-- that is, inf\/initely negative!
  Plotting the exact $E_\text{bdry}(t,x)$ for various small but nonzero
  values of  $t$ reveals a steep rise as $x$ approaches the
  boundary, followed by an even steeper plunge to negative values
  still closer to the boundary (cf.\ Fig.~\ref{fig:hfline}).
 This behavior  assures that $\int_0^L E_\text{bdry}(t,x)\,
dx=0$ for any nonzero $t$, so that the total energy \eqref{inten}
 is independent of $\xi$, as it must be.
 (The integral has been evaluated directly
  in the half-line case~\eqref{hflineint},
 and with the cutof\/f in place there is no obstacle to integrating
the original spectral sum term by term and observing that the
 total-derivative terms integrate to zero.)
 This  mathematical phenomenon was pointed out
 by Ford and Svaiter~\cite{FS}.

The situation for the odd case is very similar.
$E_\text{bdry}(t,x)$ is now an odd  function of $x-\frac L2$,
 with the same
qualitative behavior as in the even case near $x=0$ and the
inverted behavior at the other end of the interval.
 The formulas parallel to those above are
 \begin{gather*}
 E_\text{bdry}(0,x) = \frac{\pi}{8L^2} \,
 \cot\left (\frac{\pi x}{L}\right )
 \csc\left (\frac{\pi x}{L}\right ),\\
E_\text{bdry}(0,x) = \frac1{8\pi x^2} - \frac{\pi}{48L^2} +
 O(x^2),\\
E_\text{bdry}(t,x) = -\, \frac{\pi}{8L^2}
 \coth\left (\frac{\pi t}{2L}\right) \csch\left (\frac{\pi t}{2L}\right)
  +O(x^2),\\
E_\text{bdry}(t,0) \approx -\,\frac1{2\pi t^2}.
\end{gather*}

 For the problem with a single boundary at $x=0$ we have
 \begin{align*} E_\text{bdry}(t,x) &=
 -\,\frac12\,\pd{}t \int_0^\infty \sigma_\text{bdry}(\omega,x )
  e^{-\omega t}\, d\omega  = -\,\frac12\,\pd{}t \int_0^\infty \frac{(-1)^l}{\pi}\,
 \cos(2\omega x) e^{-\omega t}\, d\omega \\
 &= -\, \frac{(-1)^l}{2\pi}\, \pd{}t \,\frac{t}{t^2+4x^2}
 = \frac{(-1)^l}{2\pi}\, \frac{t^2-4x^2}{(t^2+4x^2)^2}.
 \end{align*}
Restricted to $0<x<L$,  this is exactly the $n=0$ term in the sum
for the problem with two boundaries.
 (The other short path, $n=-1$, naturally gives a symmetrical
contribution localized at the other boundary.)
 The leading terms  are the same as found above for two boundaries.
 In particular,
 \[E_\text{bdry}(0,x) = \frac{(-1)^{l-1}}{8\pi x^2},\]
 in agreement with the renormalized vacuum energy found by
 Romeo and Saharian \cite[(3.21)]{RS}.
 At small $x$ and f\/ixed $t$ we have
 \[E_\text{bdry}(t,x) = \frac{(-1)^l}{2\pi}
 \left [\frac1{t^2} - \frac{12x^2}{t^4}
 + O\left (\frac{x^4}{t^6}\right )\right ],
 \]
 again giving some insight into the sharp spike of opposite sign
that keeps the total boundary energy equal to zero as long as the
regularization has not been removed.
 Indeed, in this case the integral of the exact function is
elementary:
\begin{gather}\int_0^\infty \frac{t^2-4x^2}{(t^2 + 4x^2)^2 }\, dx
 = \left . \frac{x}{t^2+4x^2} \right |^\infty_{x=0} = 0,
 \label{hflineint}\end{gather}
 as reported in Section~\ref{sec:general}.

 \section{Conclusions}
 One of our major concerns in working out these elementary models
in complete detail has been to appraise the
  stationary-phase approximation
(which here means completely discarding the nonperiodic closed
orbits)
 and the approximation of ignoring all nonperiodic orbits except
the shortest ones
 (which inevitably leads to a comparison with the problem on the
half-line).
Let us summarize the observations, in reverse order.

 For the local vacuum energy density,
 the contribution of a short orbit is equal to the corresponding
term in the half-line problem (restricted to the interval, of
course).
 These are good approximations to the ``boundary'' part of the
exact answer near the corresponding boundary, which is the only
place where they are large.
 These boundary terms are not zero or generally small,
 so the stationary-phase approximation is not very good here.

 For the local spectral density, or its unnamed indef\/inite
integral,
 again the short orbit's contribution is equal to the half-line
expression.
 Comparison with the Dirichlet kernel again suggests that these are
good approximations near the boundary, and that these boundary terms
are not very important elsewhere~--
 although in this case they decay  by virtue of oscillation
(distributionally) more than by decrease of magnitude.

 For the total vacuum energy, globally renormalized,
 we found that the boundary terms in both the exact answer and the
half-line expression are equal to zero;
 that is, the stationary-phase approximation is exact!
 The short orbits, however, give a nonzero result;
 it is of roughly the same order of magnitude as the exact Casimir
energy coming from the periodic orbits, so it must be rejected as
wrong.
 The apparent contradiction with the local result is explained by
the dif\/ference between local and global renormalization.

 For the eigenvalue density, or its integral the counting function,
 the half-line theory and the exact theory agree (when both
endpoints are counted), and they give the correction to the
 stationary-phase theory that is necessary to take account of the
ef\/fect of the boundary conditions on the position of the lowest
eigenvalue.
 The short-orbit expression  dif\/fers, but it seems to be a
reasonable approximation.

 \subsection*{Acknowledgements}

 This work has been supported by the National Science Foundation
under Grants DMS-0405806 and PHY-0554849.

\pdfbookmark[1]{References}{ref}
 \LastPageEnding

 \end{document}